\documentclass{jfp}
                                                                                   
\usepackage{MathSpad,natbib}
                                                                                   
\begin{document}

\bibliographystyle{natbib}                                                                                  
  
 \journaltitle{}
\cpr{}
\doival{}    
                                                                             
\lefttitle{Roland Backhouse, Walter Guttmann, Michael Winter}
\righttitle{An Example of Goal-directed, Calculational Proof}
    
 \totalpg{}
\jnlDoiYr{}    
                                                                           
\begin{authgrp}
\author{Roland Backhouse, \url{https://orcid.org/0000-0002-0140-8089}}
\affiliation{School of Computer Science\\
        University of Nottingham\\Nottingham UK
        (\email{roland.backhouse@nottingham.ac.uk})}
\author{Walter Guttmann, \url{https://orcid.org/0000-0003-2969-1688}}
\affiliation{Computer Science and Software Engineering\\
        University of Canterbury\\Christchurch, New Zealand
        (\email{walter.guttmann@canterbury.ac.nz})}
\author{Michael Winter, \url{https://orcid.org/0000-0003-0847-0448}}
\affiliation{Department of Computer Science\\
        Brock University\\St.\ Catharines, ON, Canada
        (\email{mwinter@brocku.ca})}
\end{authgrp}

\title{An example of a goal-directed, calculational proof}
\begin{abstract}An equivalence relation can be constructed from a given  (homogeneous,  
binary)  relation in two steps:  first,  construct the smallest  reflexive and transitive relation  containing the 
given relation (the ``star'' of the relation)  and, second,  construct the largest symmetric relation that is
included in the result of the first step.  The  fact that  the final result is also  reflexive and transitive (as
well as symmetric), and thus an equivalence relation, 
 is not immediately obvious, although straightforward to prove.  
Rather than prove that the defining properties of reflexivity and transitivity are satisfied,  
we establish reflexivity and transitivity \emph{constructively} by exhibiting  a starth  root --- 
 in a way that emphasises the creative process in its construction.  The resulting construction is 
fundamental to algorithms that determine
the strongly connected components of a graph as well as the decomposition of a graph into its strongly
connected  components together with an acyclic graph connecting such components.\end{abstract}

\maketitle[T]

\section{Introduction}\label{BGWintro}

Given a (homogeneous, binary) relation $R$,  the relation \setms{0.15em}$R^{*}$ is the smallest reflexive and transitive
relation containing $R$,  and $R\ms{1}{\cap}\ms{1}R^{\MPrev}$ is the largest symmetric relation that is included in $R$.  By applying
these two constructions in order, the resulting  relation 
$R^{*}\ms{1}{\cap}\ms{1}(R^{*})^{\MPrev}$ is obviously symmetric;  less obvious is that it is also reflexive and transitive,   i.e.,\ the relation 
is an equivalence relation.  

\citet[theorem 139]{BDGW22} prove that,  for all relations $R$,  the 
relation $R^{*}\ms{1}{\cap}\ms{1}(R^{*})^{\MPrev}$ can be reformulated using the identity\begin{equation}\label{BGWtheorem}
R^{*}\ms{1}{\cap}\ms{1}(R^{*})^{\MPrev}\ms{5}{=}\ms{5}(R\ms{1}{\cap}\ms{1}(R^{\MPrev})^{*})^{*}~~.
\end{equation}In words,  $R\ms{1}{\cap}\ms{1}(R^{\MPrev})^{*}$ is a \emph{starth root} of the equivalence   relation $R^{*}\ms{1}{\cap}\ms{1}(R^{*})^{\MPrev}$.  
By proving the property (\ref{BGWtheorem}), one
establishes \emph{constructively} that the relation $R^{*}\ms{1}{\cap}\ms{1}(R^{*})^{\MPrev}$ is indeed reflexive and transitive, and thus an
equivalence relation.  

(In general, a starth root of a reflexive-transitive  relation $U$ is a relation $V$ such that $U\ms{1}{=}\ms{1}V^{*}$.   Since 
 $U$ is reflexive and transitive equivales  $U\ms{1}{=}\ms{1}U^{*}$, every reflexive-transitive 
relation is a starth root of itself.   The importance of a starth root  becomes evident when it has particular
properties, such as some form of minimality.)

Because of its constructive nature,  the  identity (\ref{BGWtheorem}) plays a significant role in algorithms
that  exploit the decomposition of a finite graph into an acyclic graph together with a collection of
strongly connected components.  (In this application, the relation $R$ corresponds to the edge relation on
nodes defined by the graph,  and   $R^{*}\ms{1}{\cap}\ms{1}(R^{*})^{\MPrev}$ is the relation that holds of two nodes  when they are
both in the same strongly connected component.  Readers unfamiliar with the notation and/or property
are referred to section \ref{BGWapp} for a brief summary.)  However, as observed in
\cite[]{BDGW22},  the proof left a lot to be desired since it  used the definition of the  star
operator (reflexive-transitive closure)  as a sum of powers of $R$ together with a  quite complicated 
induction property.  Attempts we had made to apply fixed-point fusion had failed.

Recently Guttmann  formulated a  proof using the inductive definition of $R^{*}$ in point-free relation algebra.
 Winter made some improvements to Guttmann's proof.    

Originally, the Guttmann--Winter proof was presented in the traditional mathematical style:  a
bottom-up proof that miraculously ends in the final step with the desired property.  In this note,  the
proof has been rewritten in a way that emphasises the heuristics that were used to construct the proof.  
Some comments on how to present difficult proofs follow the calculations.

\section{Relation Algebra}\label{BGWapp}

In the proof we use a number of properties without specific mention.  These properties will be
known to readers well-versed in relation algebra but for others may not be so.  
For this reason,  we give a very brief summary of the relevant properties.

Variables $R$, $S$ and $T$ in the proof all denote homogeneous binary relations of the same type.  The set
notation we use (``${\subseteq}$'', ``${\cap}$'' and ``${\cup}$'') has its standard meaning  and we do assume familiarity with the 
properties of the set operators.  
A property that may be less familiar is that, for all $S$,  the function ${\cap}\ms{1}S$ has
an upper adjoint,  which we denote by $S{\rightarrow}$.  That is, for all $R$, $S$ and $T$,  \begin{equation}\label{HeytingGC}
R\ms{1}{\cap}\ms{1}S\ms{2}{\subseteq}\ms{2}T\ms{6}{\equiv}\ms{6}R\ms{2}{\subseteq}\ms{2}S{\rightarrow}T~~.
\end{equation}The property is a consequence of the universal distributivity of set-intersection over set-union.
We  call it the ``Heyting Galois connection'' because it is  essentially the same as the adjunction 
 between ${\wedge}p$   and $p{\Rightarrow}$  (for all propositions $p$) in intuitionistic logic, the formalisation of which is generally
attributed to Heyting. 

Relation composition and converse are denoted by ``${\MPcomp}$'' and ``$^{\MPrev}$'', respectively, and the identity relation is
denoted by $I$.   All of intersection, union,
composition and converse are monotonic with respect to the subset ordering.   

 Converse is defined by the Galois connection, for all $R$ and $S$, \begin{displaymath}R^{\MPrev}\ms{1}{\subseteq}\ms{2}S\ms{5}{\equiv}\ms{5}R\ms{1}{\subseteq}\ms{2}S^{\MPrev}\end{displaymath}together with the distributivity property, for all $R$ and $S$, \begin{displaymath}(R{\MPcomp}S)^{\MPrev}\ms{3}{=}\ms{3}S^{\MPrev}\ms{1}{\MPcomp}\ms{1}R^{\MPrev}\end{displaymath}and the property that \begin{displaymath}I^{\MPrev}\ms{2}{=}\ms{2}I~~.\end{displaymath}The modularity rule (aka the Dedekind rule) is used in both its forms: for all $R$, $S$ and $T$,\begin{displaymath}R{\MPcomp}S\ms{2}{\cap}\ms{2}T\ms{5}{\subseteq}\ms{5}R{\MPcomp}(S\ms{3}{\cap}\ms{3}R^{\MPrev}\ms{1}{\MPcomp}\ms{1}T)\end{displaymath}and its symmetric counterpart\begin{displaymath}R\ms{2}{\cap}\ms{2}S{\MPcomp}T\ms{5}{\subseteq}\ms{5}(R\ms{1}{\MPcomp}\ms{1}T^{\MPrev}\ms{3}{\cap}\ms{3}S){\MPcomp}T~~.\end{displaymath}The rule is important because composition does not distribute over intersection:  it gives a handle on
expressions involving both operators where the intersection is on the lower side of a set inclusion.

$R^{*}$ denotes the reflexive, transitive closure of $R$.  The inductive definition of $R^{*}$ used here\footnote{An
alternative fixed-point definition ---alluded to in the text--- is the direct formalisation of the property
that $R^{*}$ is the least reflexive, transitive relation that contains $R$.} is the combination of the two properties\begin{displaymath}I\ms{3}{\cup}\ms{3}R{\MPcomp}R^{*}\ms{4}{\subseteq}\ms{4}R^{*}\end{displaymath}and, for all $T$,  \begin{displaymath}R^{*}\ms{1}{\subseteq}\ms{1}T\ms{5}{\Leftarrow}\ms{7}I\ms{1}{\cup}\ms{1}R{\MPcomp}T\ms{2}{\subseteq}\ms{2}T~~.\end{displaymath}That is, $R^{*}$ is the least prefix point of the function mapping $T$ to $I\ms{2}{\cup}\ms{2}R{\MPcomp}T$.  We don't directly  use the
fact that $R^{*}$ is a fixed point of this function but we do use the (derived) property that, for all $R$, \begin{displaymath}I\ms{2}{\subseteq}\ms{2}R^{*}\ms{8}{\wedge}\ms{8}(R^{*})^{*}\ms{3}{=}\ms{3}R^{*}\ms{8}{\wedge}\ms{8}R^{*}\ms{1}{\MPcomp}\ms{1}R^{*}\ms{4}{=}\ms{4}R^{*}\ms{8}{\wedge}\ms{8}(R^{\MPrev})^{*}\ms{2}{=}\ms{2}(R^{*})^{\MPrev}~~.\end{displaymath}We also use the fact that the star operator is monotonic with respect to the subset ordering.  

It is mentioned in the introduction that the identity we have proved is central to a number of algorithms
that exploit graph theory.  In such algorithms,  the relation $R$ is the edge relation on nodes of a finite directed
graph:  specifically, two nodes $u$ and $v$  are related by $R$ iff there is an edge in the corresponding graph  
from $u$ to $v$.  Conversely,   two nodes $u$ and $v$  are related by $R^{\MPrev}$ iff there is an edge in the graph from $v$ to $u$.
 The graph corresponding to $R^{\MPrev}$ is thus the graph obtained by reversing the edges of the graph
corresponding to $R$. Nodes $u$ and $v$ are related by $R^{*}$  iff there is a path from $u$ to $v$  in the graph, and by
$(R^{\MPrev})^{*}$ iff  there is a path  from $u$ to $v$ in the graph formed of reversed edges.  Equivalently,  $u$ and $v$ are
related by  $(R^{\MPrev})^{*}$ iff there is a path from $v$ to $u$ in the graph.  Formally, the equivalence is expressed by the
identity $(R^{\MPrev})^{*}\ms{2}{=}\ms{2}(R^{*})^{\MPrev}$. 

The relation $R^{*}\ms{1}{\cap}\ms{1}(R^{\MPrev})^{*}$ holds between nodes $u$ and $v$ iff  there is both a path from $u$ to $v$ and a path from $v$
to $u$ in the corresponding graph.  Thus $R^{*}\ms{1}{\cap}\ms{1}(R^{\MPrev})^{*}$ is the  relation that holds between nodes $u$ and $v$ when
both are in the same strongly connected component of the graph; moreover,   since  $(R^{\MPrev})^{*}\ms{2}{=}\ms{2}(R^{*})^{\MPrev}$, it 
equals  $R^{*}\ms{1}{\cap}\ms{1}(R^{*})^{\MPrev}$.  

The relation $R\ms{1}{\cap}\ms{1}(R^{\MPrev})^{*}$ holds between nodes $u$ and $v$ iff there is an edge from $u$ to $v$ and a path from $v$ to $u$.  
The  identity (\ref{BGWtheorem}) thus states that nodes $u$ and $v$ are strongly connected   iff there is a path
from $u$ to $v$ in the graph  corresponding to this relation.  This insight is fundamental to algorithms that
determine the strongly connected components of a graph as well as the decomposition of a graph into its
strongly connected  components together with an acyclic graph connecting such components.

\section{The Proof}\label{BGWproof}

As stated previously,  for arbitrary relation $R$,  the relation $R^{*}\ms{1}{\cap}\ms{1}(R^{*})^{\MPrev}$ is an equivalence relation; it is, thus, 
reflexive and transitive.  Our goal is to establish reflexivity and transitivity   \emph{constructively}
by calculating  a starth root of the relation.  That is, we aim to calculate a relation $T$  such that
 $R^{*}\ms{1}{\cap}\ms{1}(R^{*})^{\MPrev}\ms{2}{=}\ms{2}T^{*}$.

The form of this goal suggests that the fixed-point fusion theorem is applicable:  $R^{*}$ is a least  fixed
point and, for all $S$,   the function  ${\cap}S$  is a lower adjoint in a Galois connection (the Heyting connection
mentioned above).   These are precisely the circumstances in which fusion is applicable:  the theorem
provides sufficient conditions on $R$ and $S$ under which $R^{*}$ and ${\cap}S$ can be ``fused'' into a fixed point of the 
form  $T^{*}$.   However, our efforts to achieve the goal in this way failed: the conditions
required by the fusion theorem are just too strong.  

In view of this,  we are obliged to substantially weaken our goal.  The inclusion \begin{displaymath}(R\ms{1}{\cap}\ms{1}S^{*})^{*}\ms{4}{\subseteq}\ms{4}R^{*}\ms{1}{\cap}\ms{1}S^{*}~~.\end{displaymath}(for all $R$ and $S$)  is very  easily proved and, since $(R^{\MPrev})^{*}\ms{2}{=}\ms{2}(R^{*})^{\MPrev}$, it immediately follows that \begin{displaymath}(R\ms{1}{\cap}\ms{1}(R^{\MPrev})^{*})^{*}\ms{4}{\subseteq}\ms{4}R^{*}\ms{1}{\cap}\ms{1}(R^{*})^{\MPrev}~~.\end{displaymath}(The full details are given at the end of this section.)  This means that to prove (\ref{BGWtheorem}) it suffices
to prove the converse  inclusion \begin{displaymath}R^{*}\ms{1}{\cap}\ms{1}(R^{*})^{\MPrev}\ms{4}{\subseteq}\ms{4}(R\ms{1}{\cap}\ms{1}(R^{\MPrev})^{*})^{*}~~.\end{displaymath}This, in turn, suggests the  weaker goal:  we try to determine conditions on $R$, $S$ and $T$ such that \begin{displaymath}R^{*}\ms{1}{\cap}\ms{1}S\ms{4}{\subseteq}\ms{4}T^{*}~~.\end{displaymath}The calculation is  guided by the fact that the  condition on $S$  must be satisfied by $(R^{\MPrev})^{*}$ and the
condition on $T$ by  $R\ms{1}{\cap}\ms{1}(R^{\MPrev})^{*}$  but we may be lucky and find weaker conditions.  (In fact, we don't --- but it is worth a try.)

Let us begin the calculation:
\begin{mpdisplay}{0.15em}{6.5mm}{1mm}{2}
	$R^{*}\ms{1}{\cap}\ms{1}S\ms{4}{\subseteq}\ms{4}T^{*}$\push\-\\
	$=$	\>	\>$\{$	\>\+\+\+Heyting  Galois connection\-\-$~~~ \}$\pop\\
	$R^{*}\ms{4}{\subseteq}\ms{4}S\ms{1}{\rightarrow}\ms{1}T^{*}$\push\-\\
	$\Leftarrow$	\>	\>$\{$	\>\+\+\+fixed-point definition of $R^{*}$\-\-$~~~ \}$\pop\\
	$I\ms{2}{\cup}\ms{2}R{\MPcomp}(S\ms{1}{\rightarrow}\ms{1}T^{*})\ms{5}{\subseteq}\ms{5}S\ms{1}{\rightarrow}\ms{1}T^{*}$\push\-\\
	$=$	\>	\>$\{$	\>\+\+\+Heyting  Galois connection\-\-$~~~ \}$\pop\\
	$(I\ms{2}{\cup}\ms{2}R{\MPcomp}(S\ms{1}{\rightarrow}\ms{1}T^{*}))\ms{1}{\cap}\ms{1}S\ms{4}{\subseteq}\ms{4}T^{*}$\push\-\\
	$=$	\>	\>$\{$	\>\+\+\+distributivity\-\-$~~~ \}$\pop\\
	$(I\ms{1}{\cap}\ms{1}S)\ms{2}{\cup}\ms{2}(R{\MPcomp}(S\ms{1}{\rightarrow}\ms{1}T^{*})\ms{2}{\cap}\ms{2}S)\ms{5}{\subseteq}\ms{5}T^{*}$\push\-\\
	$=$	\>	\>$\{$	\>\+\+\+Galois connection defining ``${\cup}$''\-\-$~~~ \}$\pop\\
	$I\ms{1}{\cap}\ms{1}S\ms{2}{\subseteq}\ms{2}T^{*}\ms{5}{\wedge}\ms{5}R{\MPcomp}(S\ms{1}{\rightarrow}\ms{1}T^{*})\ms{2}{\cap}\ms{2}S\ms{4}{\subseteq}\ms{4}T^{*}$\push\-\\
	$=$	\>	\>$\{$	\>\+\+\+$I\ms{1}{\subseteq}\ms{1}T^{*}$ \-\-$~~~ \}$\pop\\
	$R{\MPcomp}(S\ms{1}{\rightarrow}\ms{1}T^{*})\ms{2}{\cap}\ms{2}S\ms{4}{\subseteq}\ms{4}T^{*}~~.$
\end{mpdisplay}
Summarising, we have proved that, for all $R$, $S$ and $T$,\begin{equation}\label{BGW1stcalculation}
R^{*}\ms{1}{\cap}\ms{1}S\ms{4}{\subseteq}\ms{4}T^{*}\ms{7}{\Leftarrow}\ms{7}R{\MPcomp}(S\ms{1}{\rightarrow}\ms{1}T^{*})\ms{2}{\cap}\ms{2}S\ms{4}{\subseteq}\ms{4}T^{*}~~.
\end{equation}So far the steps taken have been  relatively routine.  The next  steps are less so:  we  seek  a condition on $S$ 
that enables the elimination of ``$S{\rightarrow}$''.  To this end, we calculate:
\begin{mpdisplay}{0.15em}{6.5mm}{1mm}{2}
	$R{\MPcomp}(S\ms{1}{\rightarrow}\ms{1}T^{*})\ms{2}{\cap}\ms{2}S$\push\-\\
	$\subseteq$	\>	\>$\{$	\>\+\+\+modularity rule\-\-$~~~ \}$\pop\\
	$R\ms{1}{\MPcomp}\ms{1}(S\ms{1}{\rightarrow}\ms{1}T^{*}\ms{4}{\cap}\ms{4}R^{\MPrev}\ms{1}{\MPcomp}\ms{1}S)$\push\-\\
	$\subseteq$	\>	\>$\{$	\>\+\+\+introduce assumption as prelude to  cancellation:\\
	$\bullet~~~~ R^{\MPrev}\ms{1}{\MPcomp}\ms{1}S\ms{3}{\subseteq}\ms{3}S$\-\-$~~~ \}$\pop\\
	$R\ms{1}{\MPcomp}\ms{1}(S\ms{1}{\rightarrow}\ms{1}T^{*}\ms{4}{\cap}\ms{4}S)$\push\-\\
	$\subseteq$	\>	\>$\{$	\>\+\+\+(Heyting Galois connection) cancellation, \\
	and monotonicity of composition\-\-$~~~ \}$\pop\\
	$R{\MPcomp}T^{*}~~.$
\end{mpdisplay}
In this way, we have derived the property that,   for all $R$, $S$ and $T$,\begin{equation}\label{BGWinclusion}
R{\MPcomp}(S\ms{1}{\rightarrow}\ms{1}T^{*})\ms{2}{\cap}\ms{2}S\ms{4}{\subseteq}\ms{4}R{\MPcomp}T^{*}\ms{7}{\Leftarrow}\ms{7}R^{\MPrev}\ms{1}{\MPcomp}\ms{1}S\ms{3}{\subseteq}\ms{3}S~~.
\end{equation}Note that the condition $R^{\MPrev}\ms{1}{\MPcomp}\ms{1}S\ms{3}{\subseteq}\ms{3}S$ is indeed satisfied by $S\ms{1}{=}\ms{1}(R^{\MPrev})^{*}$.

We now continue the calculation that led to (\ref{BGW1stcalculation}).
\begin{mpdisplay}{0.15em}{6.5mm}{1mm}{2}
	$R{\MPcomp}(S\ms{1}{\rightarrow}\ms{1}T^{*})\ms{2}{\cap}\ms{2}S\ms{4}{\subseteq}\ms{4}T^{*}$\push\-\\
	$=$	\>	\>$\{$	\>\+\+\+the hardest step in the calculation: as a prelude to applying  (\ref{BGWinclusion}), \\
	we exploit the idempotency of set-intersection\-\-$~~~ \}$\pop\\
	$R{\MPcomp}(S\ms{1}{\rightarrow}\ms{1}T^{*})\ms{3}{\cap}\ms{3}S\ms{2}{\cap}\ms{2}S\ms{5}{\subseteq}\ms{5}T^{*}$\push\-\\
	$\Leftarrow$	\>	\>$\{$	\>\+\+\+(\ref{BGWinclusion}) and monotonicity\-\-$~~~ \}$\pop\\
	$R^{\MPrev}\ms{1}{\MPcomp}\ms{1}S\ms{2}{\subseteq}\ms{2}S\ms{7}{\wedge}\ms{7}R{\MPcomp}T^{*}\ms{2}{\cap}\ms{2}S\ms{4}{\subseteq}\ms{4}T^{*}$\push\-\\
	$\Leftarrow$	\>	\>$\{$	\>\+\+\+aiming for fixed-point definition of $T^{*}$,  use modularity rule\-\-$~~~ \}$\pop\\
	$R^{\MPrev}\ms{1}{\MPcomp}\ms{1}S\ms{2}{\subseteq}\ms{2}S\ms{6}{\wedge}\ms{6}(R\ms{2}{\cap}\ms{2}S\ms{1}{\MPcomp}\ms{1}(T^{*})^{\MPrev})\ms{1}{\MPcomp}\ms{1}T^{*}\ms{2}{\subseteq}\ms{2}T^{*}$\push\-\\
	$\Leftarrow$	\>	\>$\{$	\>\+\+\+fixed-point definition of $T^{*}$\-\-$~~~ \}$\pop\\
	$R^{\MPrev}\ms{1}{\MPcomp}\ms{1}S\ms{2}{\subseteq}\ms{2}S\ms{7}{\wedge}\ms{7}R\ms{2}{\cap}\ms{2}S\ms{1}{\MPcomp}\ms{1}(T^{*})^{\MPrev}\ms{3}{\subseteq}\ms{3}T$\push\-\\
	$\Leftarrow$	\>	\>$\{$	\>\+\+\+aiming for  $S\ms{2}{=}\ms{2}(R^{\MPrev})^{*}$, rewrite $(T^{*})^{\MPrev}$ as $(T^{\MPrev})^{*}$
	\\ and introduce conditions $(T^{\MPrev})^{*}\ms{1}{\subseteq}\ms{1}S$ and $S{\MPcomp}S\ms{1}{\subseteq}\ms{1}S$ \-\-$~~~ \}$\pop\\
	$R^{\MPrev}\ms{1}{\MPcomp}\ms{1}S\ms{2}{\subseteq}\ms{2}S\ms{6}{\wedge}\ms{6}(T^{\MPrev})^{*}\ms{1}{\subseteq}\ms{1}S\ms{6}{\wedge}\ms{6}S{\MPcomp}S\ms{2}{\subseteq}\ms{2}S\ms{6}{\wedge}\ms{6}R\ms{1}{\cap}\ms{1}S\ms{2}{\subseteq}\ms{2}T$\push\-\\
	$\Leftarrow$	\>	\>$\{$	\>\+\+\+1st conjunct:  fixed-point definition of $(R^{\MPrev})^{*}$\\
	4th conjunct:  reflexivity of ${\subseteq}$ \\
	3rd conjunct:  transitivity of $(R^{\MPrev})^{*}$\\
	2nd conjunct:  $R{\cap}S\ms{1}{\subseteq}\ms{1}R$  and monotonicity of converse and star\-\-$~~~ \}$\pop\\
	$S\ms{2}{=}\ms{2}(R^{\MPrev})^{*}\ms{5}{\wedge}\ms{5}T\ms{2}{=}\ms{2}R\ms{1}{\cap}\ms{1}S~~.$
\end{mpdisplay}
Summarising the calculation,  we have proved that, for all $R$, $S$ and $T$,  \begin{equation}\label{2ndcalculation}
R{\MPcomp}(S\ms{1}{\rightarrow}\ms{1}T^{*})\ms{2}{\cap}\ms{2}S\ms{4}{\subseteq}\ms{4}T^{*}\ms{7}{\Leftarrow}\ms{7}S\ms{2}{=}\ms{2}(R^{\MPrev})^{*}\ms{6}{\wedge}\ms{6}T\ms{2}{=}\ms{2}R\ms{1}{\cap}\ms{1}(R^{\MPrev})^{*}~~.
\end{equation}Combining (\ref{BGW1stcalculation}) and (\ref{2ndcalculation}),  we get \begin{equation}\label{BGWtheorem.sub}
R^{*}\ms{1}{\cap}\ms{1}(R^{\MPrev})^{*}\ms{4}{\subseteq}\ms{4}(R\ms{1}{\cap}\ms{1}(R^{\MPrev})^{*})^{*}~~.
\end{equation}As mentioned earlier,  the opposite inclusion  is easy to prove:
\begin{mpdisplay}{0.15em}{6.5mm}{1mm}{2}
	$(R\ms{1}{\cap}\ms{1}(R^{\MPrev})^{*})^{*}\ms{4}{\subseteq}\ms{4}R^{*}\ms{1}{\cap}\ms{1}(R^{\MPrev})^{*}$\push\-\\
	$=$	\>	\>$\{$	\>\+\+\+Galois connection defining intersection\-\-$~~~ \}$\pop\\
	$(R\ms{1}{\cap}\ms{1}(R^{\MPrev})^{*})^{*}\ms{3}{\subseteq}\ms{3}R^{*}\ms{7}{\wedge}\ms{7}(R\ms{1}{\cap}\ms{1}(R^{\MPrev})^{*})^{*}\ms{3}{\subseteq}\ms{3}(R^{\MPrev})^{*}$\push\-\\
	$\Leftarrow$	\>	\>$\{$	\>\+\+\+1st conjunct:  star is monotonic\\
	2nd conjunct:  $(R^{*})^{*}\ms{3}{=}\ms{3}R^{*}$ (with $R\ms{1}{:=}\ms{1}R^{\MPrev}$) and star is monotonic  \-\-$~~~ \}$\pop\\
	$R\ms{1}{\cap}\ms{1}(R^{\MPrev})^{*}\ms{4}{\subseteq}\ms{4}R\ms{8}{\wedge}\ms{8}R\ms{1}{\cap}\ms{1}(R^{\MPrev})^{*}\ms{4}{\subseteq}\ms{4}(R^{\MPrev})^{*}$\push\-\\
	$=$	\>	\>$\{$	\>\+\+\+Galois connection defining intersection\-\-$~~~ \}$\pop\\
	$R\ms{1}{\cap}\ms{1}(R^{\MPrev})^{*}\ms{4}{\subseteq}\ms{4}R\ms{1}{\cap}\ms{1}(R^{\MPrev})^{*}$\push\-\\
	$=$	\>	\>$\{$	\>\+\+\+reflexivity of ${\subseteq}$\-\-$~~~ \}$\pop\\
	$\mathsf{true}~~.$
\end{mpdisplay}
The identity  (\ref{BGWtheorem}) now follows from the anti-symmetry of the subset relation and the fact
that  $(R^{\MPrev})^{*}\ms{1}{=}\ms{1}(R^{*})^{\MPrev}$.  

\section{Specific Comments}\label{BGWcomments}

Before making more general remarks,    some comments on the calculation are in order.

The central problem  in the initial calculations is how to deal with the occurrence of the intersection
operator (``${\cap}$'')  on the lower side of an inclusion (``${\subseteq}$'').  

The first calculation is quite straightforward and relatively self-evident:  $R^{*}$ is by definition a least fixed
point and it is very common to use fixed-point induction to establish less obvious properties. (Formally,
fixed-point induction is the rule that a least fixed point is a least prefix point.  In this case, the rule used is
that,   for all $R$ and $S$,  \begin{displaymath}R^{*}\ms{1}{\subseteq}\ms{1}S\ms{5}{\Leftarrow}\ms{5}I\ms{1}{\cup}\ms{1}R{\MPcomp}S\ms{2}{\subseteq}\ms{2}S~~.\end{displaymath}There is a choice of which fixed-point definition of $R^{*}$ to use should the calculation fail.)
The combination of fixed-point induction with the use of a Galois connection is also very common.  In
this case, the ``Heyting" Galois connection is, for all $R$, $S$ and $T$,  \begin{displaymath}R\ms{1}{\cap}\ms{1}S\ms{2}{\subseteq}\ms{2}T\ms{6}{\equiv}\ms{6}R\ms{2}{\subseteq}\ms{2}S{\rightarrow}T~~.\end{displaymath}The problem of the intersection operator is resolved by simply ``shunting'' it out of
the way and then ``shunting'' it back.  
The remaining steps are relatively self-evident.  

The issue that must be resolved in the second calculation  is that  ``$S{\rightarrow}$'' has been introduced  on the left
side of an inclusion.  It is vital that this is eliminated.  The Heyting Galois connection suggests a line of
attack.  Specifically, we have the cancellation rule: for all $S$ and $T$,\begin{displaymath}(S{\rightarrow}T)\ms{1}{\cap}\ms{1}S\ms{3}{\subseteq}\ms{3}T~~.\end{displaymath}Aiming to apply cancellation,  the calculation  begins by applying the  modularity rule.  In this way,
(\ref{BGWinclusion})  is easily derived.

Undoubtedly, the hardest step of all is the first step of the third calculation: the step in which idempotency
of set intersection is applied to replace ``${\cap}S$'' by ``${\cap}\ms{2}S\ms{1}{\cap}\ms{1}S$''.   Effectively, instead of (\ref{BGWinclusion}), the
equivalent property \begin{equation}\label{BGWinclusion0}
R{\MPcomp}(S\ms{1}{\rightarrow}\ms{1}T^{*})\ms{2}{\cap}\ms{2}S\ms{5}{\subseteq}\ms{5}R{\MPcomp}T^{*}\ms{2}{\cap}\ms{2}S\ms{8}{\Leftarrow}\ms{8}R^{\MPrev}\ms{1}{\MPcomp}\ms{1}S\ms{3}{\subseteq}\ms{3}S
\end{equation}has been applied.    In fact, (\ref{BGWinclusion0})  can be further strengthened  by replacing the inclusion on
the consequent  by an equality since, for all $R$, $S$ and $U$, 
\begin{mpdisplay}{0.15em}{6.5mm}{1mm}{2}
	$R{\MPcomp}U\ms{2}{\cap}\ms{2}S\ms{4}{\subseteq}\ms{4}R{\MPcomp}(S{\rightarrow}U)\ms{2}{\cap}\ms{2}S$\push\-\\
	$\Leftarrow$	\>	\>$\{$	\>\+\+\+monotonicity of composition and intersection\-\-$~~~ \}$\pop\\
	$U\ms{3}{\subseteq}\ms{3}S{\rightarrow}U$\push\-\\
	$=$	\>	\>$\{$	\>\+\+\+Heyting Galois connection\-\-$~~~ \}$\pop\\
	$U\ms{1}{\cap}\ms{1}S\ms{3}{\subseteq}\ms{3}U$\push\-\\
	$=$	\>	\>$\{$	\>\+\+\+property of  intersection\-\-$~~~ \}$\pop\\
	$\mathsf{true}~~.$
\end{mpdisplay}
Thus, by antisymmetry of the subset ordering  together with  (\ref{BGWinclusion0}),\begin{equation}\label{BGWinclusion1}
R{\MPcomp}(S\ms{1}{\rightarrow}\ms{1}T^{*})\ms{1}{\cap}\ms{1}S\ms{5}{=}\ms{5}R{\MPcomp}T^{*}\ms{1}{\cap}\ms{1}S\ms{8}{\Leftarrow}\ms{8}R^{\MPrev}\ms{1}{\MPcomp}\ms{1}S\ms{3}{\subseteq}\ms{3}S~~.
\end{equation}Although the stronger property (\ref{BGWinclusion1}) is not  used directly,  its derivation provides a useful 
safety check:  because we
have derived an equality, we know that  simplifying  the expression  ``$R{\MPcomp}(S\ms{1}{\rightarrow}\ms{1}T^{*})\ms{1}{\cap}\ms{1}S$'' to ``$R{\MPcomp}T^{*}\ms{1}{\cap}\ms{1}S$''
does not incur any loss of information (so long as the condition $R^{\MPrev}\ms{1}{\MPcomp}\ms{1}S\ms{3}{\subseteq}\ms{3}S$ is satisfied).  %This is the raison d'\^{e}tre for the use of the idempotence of set intersection.

\section{General Comments}\label{BGWconc}

So much for the details of the calculation; now more general comments.

Since the earliest  days of the development of  ``correct-by-construction'' program design techniques, 
goal-directed reasoning has always been a central theme of  ``program calculation''.  For example,  
``programming as a goal-oriented activity'' was a specific topic in Gries's textbook ``The Science of
Programming''  \cite[chapter 14]{gries81} and broadening the theme to mathematical proofs in general
was the topic of Van Gasteren's thesis  \cite[]{netty90}.    Goal-directed reasoning is also evident
in many of Dijkstra's  ``EWD''s (available from the University of Texas) and many other publications of
the last fifty years.

In contrast, the standard mathematical style is ``bottom-up''.  That is evident from the fact that
mathematicians almost always use \emph{only-if} arguments (implication) as opposed to \emph{if} arguments
(follows-from).   In our view,  it is extremely important that the more challenging calculations are
presented in a goal-directed way, as we  have tried to do above.  It  is important because it helps to teach
the creative process underlying the  mathematics of program construction.  Of course, when a new
theory is being developed the work often proceeds in a bottom-up fashion: one identifies the more
straightforward properties and builds up to properties that are not so obvious.  But each step in the
process is an exploration.  One seeks properties of a certain type (for example, distributivity properties)
but the exact form of the properties is not known at the outset.  It is vital that we develop a style of
calculation that exposes the creative process and that we communicate this process to our students.   

Many calculations are, of course, straightforward and don't merit much  discussion.  Less interesting
calculations are ones where each step \emph{simplifies} the expression under consideration (in some sense of
the word ``simplify'').  In contrast, the
calculation above involves several \emph{complification} steps.  In particular, the step we have singled out as the
hardest of all is a complification step: idempotency is used in the derivation of (\ref{BGWinclusion}) 
to replace an expression of the form $X{\cap}S$ by
 $X{\cap}S{\cap}S$.  Idempotency is normally presented as a simplification rule whereby the number of
occurrences of the operator in question is reduced.  In order to foster creative calculation, it is also vital 
to avoid an undue bias in the presentation of equational properties; equality is after all a symmetric
operator.

In summary, what we have presented is, in our view, a
good example of a non-trivial calculation that deserves careful study.  We hope that,
in future, more effort is spent in research publications and textbooks on elucidating the process of
creative calculation.  Historically one argument against calculations in the style above  is the need to
save space.  But modern technology ---the much reduced reliance on ``hard copy''--- makes this
argument much less relevant. 

\section*{Acknowledgements}

Thanks to the anonymous referees for their detailed and helpful comments.

\section*{Declaration of Interests}

The authors report no conflict of interest.

\end{document}